\documentclass[iop]{emulateapj}

\usepackage{apjfonts}
\usepackage{natbib} 
\newcommand{\ie}{{\em i.e.}}
\newcommand{\eg}{{\em e.g.}}

\newcommand{\re}{\ensuremath{\rm{R}_e}}
\newcommand{\msol}{\ensuremath{\rm{M}_{\odot}}}
\newcommand{\sm}{\ensuremath{\rm{M}_*}}
\newcommand{\per}{\ensuremath{\!\!\times\!\!}}
\newcommand{\ED}{EDisCS}

\slugcomment{Draft}
\slugcomment{\today}

\shorttitle{Superdense galaxies in \ED\ clusters}
\shortauthors{Valentinuzzi et al.}

\begin{document}

\title{Superdense massive galaxies in the ESO Distant Cluster Survey (\ED)}

\author{T. Valentinuzzi$^1$, B.M. Poggianti$^2$, R.P. Saglia$^3$, A. Arag{\'o}n-Salamanca,$^4$, L. Simard$^5$, P. S{\'a}nchez-Bl{\'a}zquez$^{6,7}$, M. D'onofrio$^1$, A. Cava$^{6,7}$, W.J. Couch$^8$, J. Fritz$^2$, A. Moretti$^2$, B. Vulcani$^{1,2}$ }

\affiliation{$^1$Astronomical Department, University of Padova, Italy,
$^2$INAF-Astronomical Observatory of Padova, Italy, 
$^3$Max-Planck Institut f{\"u}r Extraterrestrische Physik, Giessenbachstra{\ss}e, D-85741 Garching, Germany,
$^4$School of Physics and Astronomy, University of Nottingham, University Park, Nottingham NG7 2RD, United Kingdom,
$^5$Herzberg Institute of Astrophysics, National Research Council of Canada, Victoria, BC V9E 2E7, Canada,
$^6$Instituto de Astrof{\'i}sica de Canarias, E-38200 La Laguna, Tenerife, Spain,
$^7$Departamento de Astrof{\'i}sica, Universidad de La Laguna, E-38205 La Laguna, Tenerife, Spain,
$^8$Center for Astrophysics and Supercomputing, Swinburne University of Technology, Australia,
}

\begin{abstract} 
We find a significant number of massive and compact galaxies in clusters from the ESO Distant Clusters Survey (\ED) at
$0.4<z<1$.  They have similar stellar masses, ages, sizes and axial
ratios to local $z\sim0.04$ compact galaxies in WINGS clusters,
and to $z=1.4-2$ massive and passive galaxies found in the general
field.  If non-BCG cluster galaxies of all densities, morphologies and spectral types are considered, the
median size of \ED\ galaxies is only a factor 1.18 smaller than in
WINGS. We show that for morphologically selected samples,
the morphological evolution taking place in a significant
fraction of galaxies during the last Gyrs may introduce an apparent,
spurious evolution of size with redshift, which is actually due to 
intrinsic differences in the selected samples.  We conclude that the
median mass-size relation of cluster galaxies does not evolve
significantly from $z\sim0.7$ to $z\sim0.04$. In contrast, the masses and sizes of BCGs and galaxies with $\sm\!\!>4\per10^{11}\msol$ have significantly increased by a factor of 2 and 4, respectively, confirming the results of a number of recent works on the subject.
Our findings show that progenitor bias effects play an important role in the size-growth paradigm of massive and passive galaxies. 
\end{abstract}


\maketitle

\section{Introduction} 
High-z studies (as far as $z\sim2.4$ ) have found a significant number of massive, 
passively evolving galaxies
(stellar mass $\sm>10^{10}\rm{M_{\odot}}$) with relatively small effective radii $\re<2\rm{kpc}$
\citep[see, among
others,][]{trujillo06,cimatti08,vandokkum08,vanderwel09,saracco09}, 
sometimes named superdense\footnote{ Regarding \emph{physical densities}, these galaxies are anyway thought not to be extreme 
\citep[see, \eg,][]{bezanson09,hopkins09a}} galaxies (SDGs). The general claim by various
authors is that local galaxies are three to six times larger in size
when compared to high-z ones, at the same stellar mass.
In addition, \citet{trujillo09} found a complete absence of massive, old and extremely compact galaxies in the local universe.

However, \citet{valentinuzzi10} (hereafter V10) have shown that 22\% of local cluster members in the WINGS sample with $\sm>3\per10^{10}\msol$ and $\Sigma_{50}\geq3\per10^{9}\msol kpc^{-2}$ have the same characteristics of the high-z SDGs reported in the literature by various authors. In the same paper, the authors found that selecting galaxies with old stellar populations
is equivalent to selecting the smaller ones, for a given stellar mass. 
Since a large number of galaxies have stopped forming stars at relatively low redshift ($z<1.4$),
and these tend to be the largest,
it is not valid to compare high-z passive galaxies with all low-z passive
ones.
To avoid selection effects when making comparisons with passive galaxies at high redshift,
one needs to select locally those galaxies which
{\it were already passive} at the cosmic time the high-z data correspond to.

More recently, \citet{taylor09} revisited the search of SDGs in
SDSS-DR7 and found a relatively
small but significant number of SDGs. Following the same criterion used in V10, they find a 1.3\% fraction of SDGs.

The issue is much debated. \citet{mancini09} have analyzed a sample
of 12 galaxies at $0.5<z<1.9$ in the Cosmos field, finding masses and
sizes compatible with the local SDSS ones. Furthermore, by using a set of simulated early-type galaxies, they have shown that the low
signal-to-noise of high-z images can cause measured effective-radii to
be lower than the intrinsic values. 
In a recent paper \citet{vandokkum10}
select galaxies with a constant number density at different
cosmic times. They use all galaxies instead of only passive ones, 
and find that galaxies have grown in size by a factor of 4 from $z\sim2$ to $z\sim0$.

Even more recently, while \citet{szomoru10}  confirm the extreme compactness of a $z=1.9$ galaxy with the HST-WFC3, \citet{saracco10} show that the comoving number density
of compact ETGs over the volume of about $4.4\per10^5\rm{Mpc}^3$ sampled by the GOODS area between $0.9<\rm{z}< 1.92$ is compatible even with the local lower limits given in V10.

In this Letter we present the results of a search for SDGs in the ESO
Distant Clusters Survey (\ED) at $z\sim0.7$, and we report the
comparison of the mass-size relation (MSR) with the same relation in
WINGS clusters at $z\sim0$. We further discuss selection effects
which may introduce a spurious size evolution with redshift if not
properly taken into account.

\section{The data}

The high-z cluster sample is extracted from
\ED, a multiwavelength photometric and spectroscopic survey of
galaxies in 20 fields containing galaxy clusters at $0.4<z<1$
\citep[][]{white05}.  We will use a sub-sample of 8 clusters\footnote{Their 
short names found in other \ED's papers are CL1138,CL1138a, CL1040, CL1216, CL1054-11, CL1054-12, CL1232, CL1354.} which
have HST-ACS images for high-precision size measurements
\citep[][]{desai07}, and cluster central velocity dispersions ($\sigma_{clus}\geq400\rm{km\;s^{-1}}$, $<\sigma_{clus}>\sim700\rm{km\;s^{-1}}$) similar to local WINGS clusters \citep[][]{halliday04,milvang08,desai07}. Three of these clusters have $z\sim0.5$, the rest of them have $z= 0.7-0.8$.

Galaxy stellar masses were estimated using the {\it kcorrect} tool
\citep[]{blanton07}\footnote{http://cosmo.nyu.edu/mb144/kcorrect/},
that models the available observed broad band photometry 
(VRIJK, or BVIJK), 
fitting templates obtained with spectrophotometric models. 
The stellar masses are defined as the mass locked into stars, including stellar remnants, at any
time, using a \citet{kroupa01} initial mass function (type 2 mass in
V10). Taking into account the statistical errors on the mass
estimates, the error of the stellar mass on individual galaxies is of the order
$\sim0.1\rm{dex}$, even though it has to be taken into account that
the scatter (rms) in the relation between masses computed with
different models is typically
$\sim0.2\rm{dex}$ \citep[for further details,
see][]{fritz07,longhetti08,vulcani10}

We use visual morphological classifications from \citet{desai07}.

We measure galaxy effective-radii \re\ with the GIM2D tool \citep[][]{simard02}
on the HST images in F814W band, by using a single component Sersic fit. The circularized \re\ is determined by numerically integrating the curve of growth of
the fitted Sersic model, and solving the equation $\rm{Flux}(\leq\re)=0.5\cdot\rm{Flux(\infty)}$
\citep[for further details, see][submitted]{saglia10}. The typical random error on the \ED's sizes is of the order of 20\% \citep[][]{simard09}.

We use a mass limited sample of \ED\ spectroscopically confirmed 
cluster members, with stellar masses~$\geq4\per10^{10}\msol$.
This mass limit corresponds to the
mass of an object whose observed magnitude is equal to the faint
magnitude limit of the spectroscopic survey, with the reddest possible color.
We correct for spectroscopic incompleteness using \citet{milvang08} 
completeness functions.

The local sample examined in this Letter comes from the WIde-field
Nearby Galaxy-clusters Survey
\citep[][]{wingsI}. WINGS\footnote{\texttt{http://web.oapd.inaf.it/wings} } is a
multiwavelength photometric and spectroscopic survey designed to
provide a robust characterization of the properties of galaxies in nearby clusters.

We use only cluster members
of the subset of WINGS clusters that have an average spectroscopic
completeness larger than 50\% (21 out of 78 clusters), and correct for spectroscopic incompleteness using the prescriptions given in \citet{wingsspe}. These WINGS clusters have redshifts $0.04<z<0.07$ and
central velocity dispersions $558<\sigma_{clus}/\rm{km\;s^{-1}}<1368$.

WINGS effective-radii, axial ratios and Sersic indexes are measured on
the V-band images with GASPHOT \citep[][]{gasphot}, an
automated tool which performs a simultaneous fit of the major and
minor axis light growth curves with a 2D flattened Sersic-law,
convolved by the appropriate, space-varying PSF. As a measure of
galaxy size we use the circularized effective radii, calculated in the
same way it was done for \ED\ sizes (see above). We note here that SDG fractions and number densities are updated accordingly in this Letter 
compared to V10 (see next sections). 
The maximum error on WINGS sizes,
based on extensive simulation runs, is of the order of 10\%
\citep[see,][]{gasphot}.

As a consistency check on sizes, we run GIM2D on one representative
V-band WINGS cluster image, to compare the resulting circularized
$R_e$ and Sersic index $n$ of $\sim800$ galaxies with GASPHOT values. We found a systematic difference in sizes
of $0.033\pm0.002\rm{dex}$, in the sense that GASPHOT sizes are larger
than GIM2D ones. This difference becomes larger (as far as
$\sim0.3\rm{dex}$) for larger galaxies, somehow confirming that GIM2D
has the tendency to systematically underestimate the sizes of the
largest galaxies at all luminosities \citep[see][]{simard02}. On the other hand, we do not find any systematic difference regarding the Sersic index estimate.

Stellar masses of WINGS galaxies have been determined by fitting the
optical spectrum (in the range $\sim 3600 \div \sim 7000$ \AA), with
the spectro-photometric model fully described in \citet{fritz07},
and correcting for color gradients outside of the fiber (see V10). 
The model derives the integrated spectrum as the combination of
stellar populations of 13 different ages, allowing dust extinction to
vary with the stellar population age and using the single metallicity
(either Z=0.05, 0.02 or 0.004) that gives the lowest $\chi^2$ fit of the observed spectrum. Although the masses were calculated in two different ways we have shown in V10 (and soon in Fritz et al. 2010 in prep.) that there is no significant systematic offset between different methods that could be capable of biasing our results.

WINGS morphologies are derived from V images using the purposely
devised tool MORPHOT. We have verified that the differences in classification between MORPHOT and an experienced human classifier are comparable to the differences between two experienced human classifiers \citep[][in preparation]{morphot}.

For the  sake of comparing the  median sizes of the two   surveys, we divide  the total sample into four mass intervals, selected to have a statistically significant number of objects in each one of them:
\begin{itemize}
\item BIN1: $4\per10^{10}\leq\sm/\msol\!<6\per10^{10}$
\item BIN2: $6\per10^{10}\leq\sm/\msol\!<1\per10^{11}$
\item BIN3: $1\per10^{11}\leq\sm/\msol\!<2\per10^{11}$
\item BIN4: $2\per10^{11}\leq\sm/\msol\leq4\per10^{11}$
\end{itemize} 
and will refer to them with the label BIN[1-4].

\section{\ED\ Superdense Galaxies}

\begin{figure*}
\centering
\includegraphics[scale=0.7]{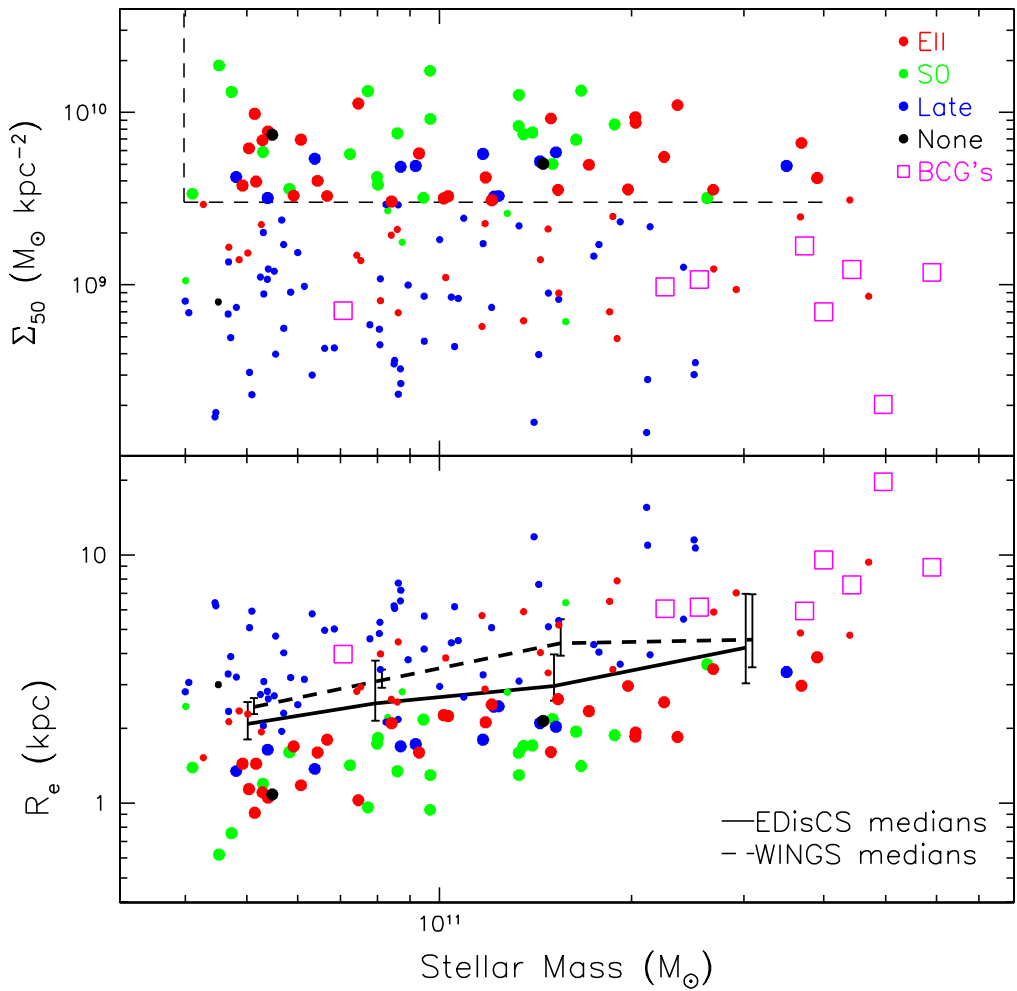}
\caption{Circularized effective radius $R_e$ and mass-density inside
  $R_e$ as a function of stellar mass for all \ED\ spectroscopic
  members galaxies with $M_*\geq4\per10^{10}\msol$.  The different
  colors mark the morphological type: blue for late-types (later than
  S0s), green for S0s, red for ellipticals and black for galaxies
  without a classification.  Bigger dots highlight the
  SDGs.  Big open magenta squares are the BCGs.  The solid and dashed
  lines in the bottom panel
  are the median completeness weighted MSRs of \ED\ and WINGS,
  respectively, obtained excluding the BCGs. Error bars are lower and
  upper quartiles of the medians. The WINGS mass medians are shifted
  by 0.01dex in X to avoid overlapping.  \label{fig:all}}
\end{figure*}

In Figure \ref{fig:all} we present the MSR (mass-size relation, bottom
panel) and the mass-density relation (top panel) of \ED\ 
cluster members with $\sm>4\per10^{10}\msol$. Colors differentiate the
morphological types (see caption and legend), large open squares are
the brightest cluster galaxies (BCGs) listed in \citet{white05}.

In the top panel, the dashed line isolates the \ED\ SDGs (larger dots
in both panels) with the same density selection criteria
($\Sigma_{50}\geq3\per10^{9}\msol kpc^{-2}$) used in V10, above the
mass completeness limit of this Letter.  These criteria were chosen to
select galaxies with mass and density ranges similar to those of
high-z ($z>1.4$) passively evolving galaxies.

As apparent in Figure \ref{fig:all}, we do find a significant number
of SDGs in the \ED\ sample. Indeed, \ED\ SDGs
represent 41\% of the total cluster population of galaxies more
massive than $\sm>4\per10^{10}\msol$.  This is an even larger
fraction than V10 found in WINGS local clusters, where 17\% are SDGs
for the mass limits and radii adopted in this Letter.  A decline with
time of the SDG fraction in clusters might be expected given that a)
the ``oldest'' galaxies in the Universe (those who stopped forming
stars very early on) inhabit clusters since very high redshifts, and
clusters accrete throughout their history galaxies with more extended
star formation histories, and b) as shown in V10, at any given mass
the oldest galaxies tend to be the most compact.  Therefore the
original population of old and compact galaxies in clusters get
progressively diluted by larger galaxies infalling into clusters at
later times.

Of the EDisCS SDGs, 41\% are ellipticals, 36\% are S0s, 20\% are
late-type galaxies, and for 4\% of them it was not possible to
assign a reliable visual morphological classification. In Table
\ref{tab:num} we present the main mean properties of \ED\ SDGs.

\begin{table} \begin{center} \caption{Completeness corrected quantities of \ED\ and WINGS SDGs. Errors on the medians are reported too. \label{tab:num}} 
\begin{tabular}{ccc}
\hline Quantity & EDisCS & WINGS\\ \hline \hline SDG fraction & 41\% &
17\% \\ Ellipticals & 41\% & 28\% \\ S0s & 36\% & 64\% \\ Late-type &
20\% & 8\% \\ Unknown morph.  & 3\% & - \\
Eff. radius $\langle\re\rangle$ & $1.70\pm0.08$ & $1.79\pm0.04$ \\ 
Sersic index $\langle n\rangle$ & $3.71\pm0.14$ & $3.21\pm0.09$ \\ 
Axial ratio $\langle b/a\rangle$ & $0.59\pm0.11$ & $0.62\pm0.03$ \\ 
Stellar mass $\langle\sm\rangle$ & $(1.08\pm0.08)\per10^{11}\msol$ & $(1.02\pm0.04)\per10^{11}\msol$ \\ 
 
\hline \hline 
\end{tabular} 
\end{center} 
\end{table}

\begin{table}
\begin{center}
\caption{Ratios of median WINGS/\ED\ sizes for the different mass intervals (see text). Errors come from the standard error propagation technique. 
\label{tab:fac}}
\begin{tabular}{ccccc}
\hline
\hline 
 WINGS/\ED      & BIN1 & BIN2 & BIN3 & BIN4 \\
All galaxies  	& $1.16^{+0.23}_{-0.17}$ &  $1.24^{+0.48}_{-0.21}$  & $1.48^{+0.56}_{-0.27}$  & $1.07^{+0.93}_{-0.59}$ \\
Early-type galaxies     & $1.76^{+0.19}_{-0.20}$ &  $1.79^{+0.44}_{-0.20}$  & $1.62^{+0.67}_{-0.36}$  & $1.28^{+0.95}_{-0.88}$ \\
WINGS Early/\ED\ All       & $1.13^{+0.22}_{-0.17}$ &  $1.18^{+0.48}_{-0.22}$  & $1.40^{+0.57}_{-0.28}$  & $1.01^{+0.98}_{-0.58}$ \\
\hline
\hline 

\end{tabular}
\end{center}
\end{table}

\begin{figure*}
\centering
\includegraphics[scale=0.7]{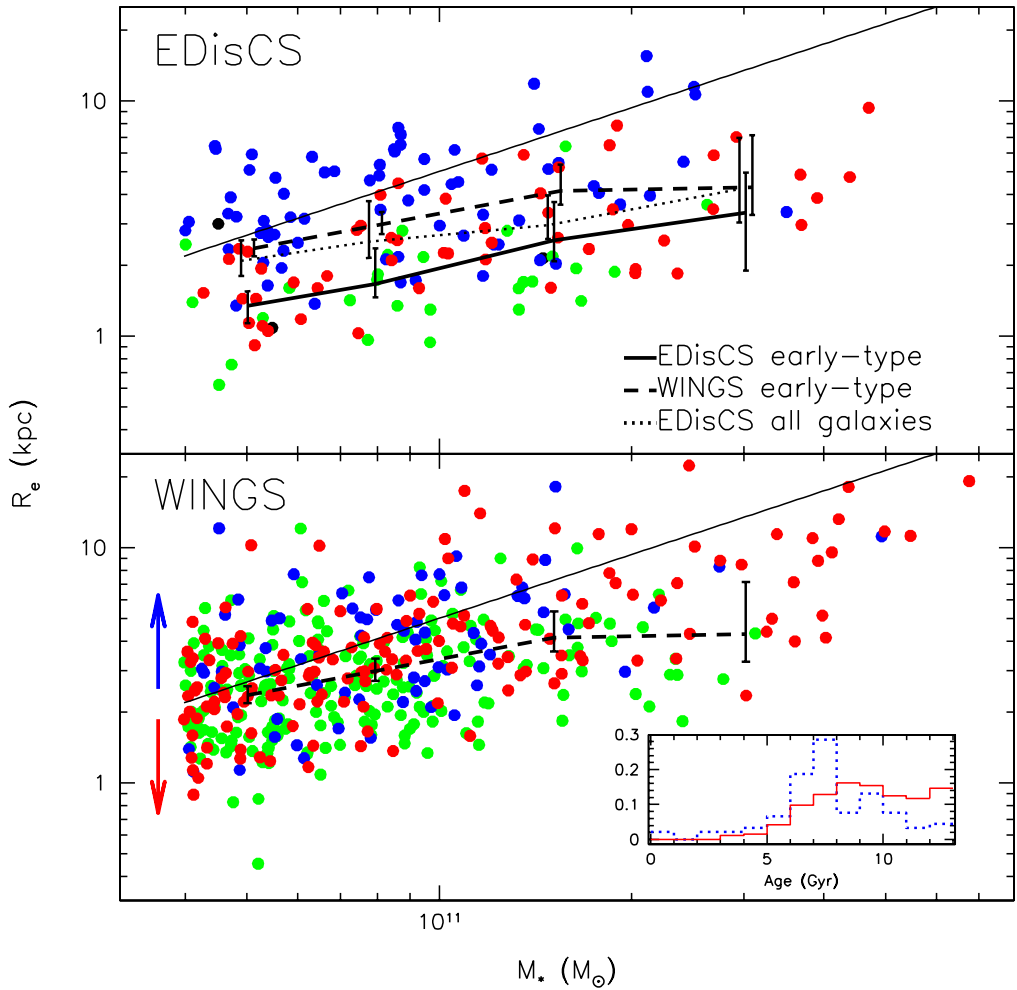}
\caption{Comparison of the mass-size relation of \ED\ (top panel)
  and WINGS (bottom panel).  Color coding is the same of Figure
  \ref{fig:all}. The black straight line delimits the area above which
there are no
  early-type galaxies in \ED.  The histogram at the
  bottom-right represents the luminosity-weighted age
  distribution of WINGS early-type galaxies above (blu dashed
  line) and below (solid red line) this line.  The black solid and
  dotted (shifted by -0.01dex in mass) lines are the median MSR for
  early-type and all types of galaxies in EDisCS, respectively. The dashed
  black line is the median MSR for WINGS early-type galaxies (shifted
  by +0.01dex). Error bars are lower and upper errors on the
  medians.\label{fig:morph}}
\end{figure*}

In the bottom panel of Figure \ref{fig:all} the median MSR of \ED\
galaxies is presented as a black solid line, and compared to the
dashed WINGS one. As we did in V10 we excluded BCGs and galaxies with
$\sm>4\per10^{11}\msol$, that are discussed separately below. 
We find that the median \re\ of \ED\ cluster galaxies with
$4\per10^{10}\leq\sm/\msol\leq4\per10^{11}$ is only a
factor 1.18 lower than the WINGS one. 
Considering separate mass bins, the maximum amount
of evolution is 1.48 (BIN3, see Table \ref{tab:fac}), while
in the other mass bins the median sizes turn out to be in good
agreement. 

These numbers may be compared with field studies that are including all morphological and spectral galaxy types. Recently, \citet{williams10} report a size evolution for {\em all galaxies} from $z\sim0.7$ to $z\sim0.04$ of  a factor of $\sim1.4$. In this regard, we stress that it is hard to draw conclusions from this comparison because it is still unclear how the incidence of SDGs and the evolution of galaxy sizes depend on environment \citep[see,][]{maltby10,rettura10}.

So far, we have seen that a considerable fraction of \ED\ cluster members are SDGs and that galaxy sizes in \ED\ and WINGS, at
all mass ranges considered, are rather similar, and do not suggest a
strong increase in size with redshift.

The BCGs and the most massive cluster galaxies with~$\sm\!\!>4\per10^{11}\msol$ have to be discussed separately, due to
their peculiar nature and evolution \citep[see, amongst others,][]{fasano10}. Indeed, the \ED\ BCGs have mean mass and size of
$\sm\sim4\per10^{11}\msol$ and $\re\sim8.5\rm{kpc}$, respectively. In
contrast, WINGS BCGs have mean values of $\sm\sim10^{12}\msol$ and
$\re\sim33.6\rm{kpc}$, suggesting that the mean size and mass of BCGs
have increased by a factor of $\sim4$ and $\sim2$ between $z\sim0.7$
and $z\sim0.04$, respectively.
Although this result seems in contrast
with \citet{whiley08}, we note that the mass of local BCGs in
that paper was calculated inside an aperture of 37kpc, which is
approximately the median half luminosity circularized radius in V band
of our local sample of BCGs. This is consistent with a picture where the
BCG progenitors increase their mass via minor mergers in the
outer regions, leaving practically unchanged the dense core
\citep[see][]{hopkins10}.  We also note that the size and mass evolution of our sample of high-z BCGs with redshift is compatible with the observational study of \citet{bernardi09} and with the theoretical expectations of \citet{delucia07}, that predict a factor of 3-4 growth in mass between $z \sim 1$ and $z=0$.

\section{Selection effects}

We have seen that the morphological fractions among the \ED\ SDGs are
considerably different from WINGS SDGs. The latter show a
larger fraction of S0s and a corresponding lower fraction of later
types. This is expected, as many studies have come
to the conclusion that 
a large fraction of today's
passive early-type galaxies have evolved from star forming late-type
galaxy progenitors in clusters \citep[][]{dressler97,fasano00,postman05,poggianti09}.

In V10 we have shown that selecting the oldest cluster galaxies means 
selecting the smallest in size. In the following, we will
highlight the biases that can be introduced by selecting galaxies
morphologically, and thus the 
importance of properly taking into account the
morphological change too. Although the morphological
evolution is strictly linked to the evolution in star formation
activity \citep[][]{poggianti09}, the time scales can be largely different \citep[][]{poggianti99,patricia09}
and thus become important at different cosmic times, in a way difficult to predict.

In Figure \ref{fig:morph} we compare \ED\ (top panel) and WINGS
(bottom panel) MSRs. Color coding is the same as for Figure \ref{fig:all}.
The black solid and dashed lines show the median MSR for early-type
galaxies in \ED\ and WINGS, respectively. Error
bars are errors on the medians. The median size of WINGS early-type
galaxies is a factor 1.53 larger than \ED's early-types, reaching an offset
as large as 1.7 at the lowest masses (BIN1 and BIN2, see Table \ref{tab:fac}).

At face value, this could be interpreted as an evolution in the sizes
of individual early-type galaxies. However, we note that the largest
\ED\ cluster members tend to 
have late-type morphology (some of which are star-forming, $\sim 70\%$, and some
passively evolving, $\sim 30\%$). 
We have arbitrarily identified a region in the
mass-size diagram, above the tilted line drawn in both panels,
where \ED\ galaxies are large and all have late-type morphologies.
In contrast, WINGS galaxies in this region are mostly (72\%) early-types,
consistent with a morphological evolution.

A convincing test of this picture is presented in the lower right
inset of Figure \ref{fig:morph}, where the distributions of the
luminosity-weighted ages of WINGS early-type galaxies above (blue
dashed) and below (red solid) the tilted line is shown. The blue-dotted
histogram is visibly sharply peaked toward lower ages, when compared
to the red one.  This is consistent with a significant fraction of EDisCS large, late-type galaxies having turned into large, passive,
early-types by $z\sim0$. 

Let's now focus only on late-types turning in S0s. Practically all \ED\ S0s are SDGs (see Figure \ref{fig:all}) and most of them are old (81\% have Luminosity-weighted-age $>3$ Gyr, 52\% $>5$ Gyr); in WINGS, instead, a large number of S0s are {\it not} SDGs (see Figure \ref{fig:morph}).
On the other hand, among the WINGS SDGs with S0 morphology, only 20\% have ages lower than the corresponding \ED\ lookback time, \ie\ were most likely morphologically changed at redshifts lower than \ED.
This is an indication that for the largest galaxies the majority of the morphological transformations took place a few billion years ago, while for most of the compact galaxies both the quenching of star formation and the final morphological type were reached at earlier epochs.

It is clear that when comparing high- with low-z samples, it is of
paramount importance to keep in mind that morphologically selecting
galaxies at different epochs introduces an apparent, but spurious size
evolution with redshift, which instead is a selection effect. Although
it is impossible\footnote{Because stellar ages cannot
be used as a proxy for morphology, given that the timescale for
morphological transformation is longer than the timescale for star formation quenching.}  from the WINGS
data to recover which galaxies were early-types at the \ED's epoch,
our findings support the hypothesis that the main reason why the
median size of WINGS early-type galaxies (dashed line in Figure
\ref{fig:morph}) is much more consistent with the median size of all \ED\
galaxies (dotted line, see also Table \ref{tab:fac}) than with
the size of only EDisCS early-types is that the
largest late-type \ED's galaxies have gradually become earlier types
by the WINGS epoch.

\section{Discussion and conclusion}

We have found that 41\% of \ED\ galaxies with
$\sm\sim4\per10^{10}\msol$ are SDGs. Their properties are similar to
WINGS SDGs, apart for a significantly different morphological
mix: the prevalence of S0s in WINGS is not found in \ED.

Such a result is not unexpected, given our previous findings: in V10
we have found that 17\% (for the mass limits and radii adopted here) of WINGS clusters members at $z \sim 0$
are SDGs. More than
50\% of them have stellar ages older than 9Gyr, a clear indication
that they were already old and compact at the \ED's epoch. 
The evolution of the SDG fraction in clusters with redshift 
is expected if
SDGs are massive and old galaxies, formed in clusters seeds and preferentially
found in today's massive clusters, while they are rarer
 in the field \citep[see][]{taylor09} and therefore in the
population of galaxies infalling into clusters at later and later times. 

We find that when galaxies of all morphological types are considered,
the median size of cluster galaxies at $z \sim 0.7$ is only a factor 1.18 smaller than the local median. 
We conclude that from $z\sim0.7$ 
to $z\sim0.04$, there is at most a very modest evolution in galaxy sizes
in clusters.

Similarly to our V10 analysis of age selection effects, we have shown that
comparing high-z morphologically selected samples with local ones can
be misleading. 
In agreement with previous results regarding the morphological evolution
in clusters, we have found that
the largest \ED\ late-type galaxies are found to be large early-types in
WINGS clusters, as it is apparent studying the morphologies above the
tilted line in Figure \ref{fig:morph}.The BCGs, instead, have been found to
evolve both in mass (a factor of $\sim2$) and size (a factor of
$\sim4$), in agreement with other recent theoretical and
observational results.

Our findings show that the progenitor bias (in age or morphology) plays an important role in the size-growth paradigm, and must be carefully taken into account when comparing local galaxy sizes with those of massive high-z galaxies.

\end{document}